%% file: paper_mue_v14.tex
\documentclass[conference]{IEEEtran}

\IEEEoverridecommandlockouts
\makeatletter
\def\markboth#1#2{\def\leftmark{\@IEEEcompsoconly{\sffamily}\MakeUppercase{\protect#1}}%
\def\rightmark{\@IEEEcompsoconly{\sffamily}\MakeUppercase{\protect#2}}}
\makeatother

\usepackage[acronyms,nonumberlist,nopostdot,nomain,nogroupskip]{glossaries}
\input{acronyms.tex}

\usepackage{kantlipsum}
\usepackage{setspace}

\usepackage[english]{babel}
\usepackage{xcolor}
\usepackage{lipsum}
\usepackage{caption}
\usepackage{cite}
\usepackage[pdftex]{graphicx}
\usepackage{subcaption}
\usepackage{amsmath}
\usepackage{booktabs}
\usepackage{colortbl}
\usepackage{amsfonts}
\usepackage{amssymb}
\usepackage{amsthm}
\usepackage{array}
\usepackage{verbatim}
\usepackage{listings}
\usepackage{algorithm}
\usepackage{algpseudocode}
\usepackage{url}
\usepackage{enumerate}
\usepackage{multirow}
\usepackage{xfrac}

\definecolor{LightBlue}{rgb}{0.5,0.5,1}
\definecolor{LightRed}{rgb}{1,0.5,0.5}
\definecolor{LightYellow}{rgb}{1,0.85,0}

\usepackage{epsfig}
\usepackage{epstopdf}
\usepackage{multicol}
\usepackage[font=footnotesize]{caption}

\usepackage{algorithm}

\makeatletter
\def\BState{\State\hskip-\ALG@thistlm}
\makeatother

\renewcommand{\arraystretch}{2}

\def\3G{\mathsf{3GPP}}

\title{Performance Assessment of MIMO Precoding on Realistic mmWave Channels}
\author{{{\textbf{$^*$Mattia Rebato}, \textbf{\IEEEauthorrefmark{2}Luca Rose}, \textbf{$^*$Michele Zorzi}} }\\ \normalsize $^*$Department of Information Engineering, University of Padova, 35131 Padova, Italy \\ \normalsize \IEEEauthorrefmark{2}Nokia Bell Labs, Paris -- Saclay, France \\ \small{$\{$\texttt{rebatoma, zorzi}$\}$\texttt{@dei.unipd.it}} -- {\texttt{luca.rose@nokia-bell-labs.com}}}

\makeglossaries
\begin{document}
\maketitle
\vspace{-0.5cm}
\captionsetup{belowskip=-2pt}
\thispagestyle{empty}
\glsunset{5g}
\glsunset{3gpp}
\glsunset{sinr}
\begin{abstract}
In this paper, the performance of multi-user \gls{mimo} systems is evaluated in terms of \gls{sinr} and capacity.
We focus on the case of a downlink single-cell scenario where different precoders have been studied.
Among the considered precoders, we range from different \gls{gob} optimization approaches to linear precoders (e.g., matched filtering and zero forcing).
This performance evaluation includes imperfect channel estimation, and is carried out over two realistic mmWave \gls{5g} propagation channels, which are simulated following either the measurement campaign done by \gls{nyu} or the \gls{3gpp} channel model.
Our evaluation allows grasping knowledge on the precoding performance in mmWave realistic scenarios. 
The results highlight the good performance of \gls{gob} optimization approaches when a realistic channel model with directionality is adopted.
\end{abstract}

\begin{IEEEkeywords}
Millimeter-wave, multi-user MIMO, 5G, interference optimization, linear precoder, grid of beams.
\end{IEEEkeywords}

\glsresetall
\glsunset{nr}
\glsunset{5g}
\glsunset{3gpp}

\vspace{-0.3cm}
\section{Introduction}
\label{introduction}
The volume of mobile data  is continuously increasing, especially with  high capacity applications that are emerging together with the next generation (i.e., \gls{5g}) of cellular communications~\cite{cisco}.
As an enabler for these capacity-intensive applications, the \gls{mmwave} band (approximately between $10$ and $300$~GHz) has been identified as a promising candidate for future mobile communications~\cite{rangan14}.
In addition to the use of \gls{mmwave} frequencies, another major aspect of the new mobile generation is the densification of the network applying small cells in large numbers.
Furthermore, \gls{mu} massive \gls{mimo} systems became of high interest as they contribute to reaching the \gls{5g} high demands (e.g., in terms of rates and densities), due to their ability to greatly increase network capacity~\cite{larsson14}. 
For this reason, it is important to study and evaluate \gls{mu} massive \gls{mimo} systems over \gls{5g} \gls{mmwave} propagation channels.
By exploiting such technologies, data transmission rates are expected to increase in the \gls{ran}, and a more efficient use of the radio spectrum can be achieved.

The purpose of \gls{mu} \gls{mimo} systems is to account for channel scattering and reflections, thus exploiting the spatial dimension and creating multiple beams of the signal in the direction of the \glspl{ue}, so that each user can benefit from the whole allowed bandwidth at any time instant.
This can be achieved by precoding the information at the \gls{gnb} side.
Using a precoder, data is distributed on the different antenna elements of the \gls{gnb} in order to perform beamforming of information toward the served \glspl{ue}.

Many works in the literature focused on the evaluation of precoding techniques for \gls{mu} \gls{mmwave} systems with massive \gls{mimo}.
The closest works to ours are~\cite{marzetta13,Hoydis13,colon15,Gao11}. 
In~\cite{marzetta13}, massive \gls{mimo} was proposed and studied under the ideal condition of almost infinite antennas. 
In~\cite{Hoydis13}, precoding techniques such as \gls{mmse}, \gls{mf} and \gls{zf} were studied under the assumption of a Rayleigh channel model and under the condition of perfect \gls{csi} acquisition. 
In~\cite{colon15}, channel estimation errors were introduced to estimate the implementation loss in terms of precoding gain, whereas in ~\cite{Gao11} the authors link the precoding performance with channel correlation.
Finally, a recent piece of work~\cite{related_work1} uses a realistic channel model to perform an evaluation of a \gls{mu} system in terms of bit error rate as a function of the number of antenna elements used at the transmitter side, while however overlooking the effect of different precoding strategies and channel estimation errors.

From the literature, it emerges that linear precoding schemes can be used to reach high performance under ideal assumptions.
Less known is however their performance when realistic channel models are considered.
To be precise, under a Rayleigh fading model, it is known that \gls{mmse} performs appreciably better in terms of balancing the resources among the \glspl{ue} acting as a trade-off between \gls{mf} and \gls{zf} approaches.
However, the Rayleigh fading model oversimplifies the channel characterization, resulting in a channel model that does not reflect the real \gls{mmwave} propagation specifics.

In \gls{3gpp} \gls{nr} systems, the exploitation of \gls{mmwave} frequency bands (both at $28$~GHz and at $60$~GHz) for the next generation of mobile communications is currently defined~\cite{38104}.
Within the standard, different types of \gls{csi} feedback mechanisms have been included to support \gls{mimo} transmissions.
In particular, release~$15$ includes Type-I and Type-II codebook \gls{csi} feedback, enabling different trade-offs between \gls{csi} resolutions and feedback overhead~\cite{38211}. 
%
More precisely, when a Type-I \gls{csi} feedback scheme is adopted, the UE feeds back the index of a vector taken from a suitable oversampled DFT codebook that best approximates the dominant eigenvector of the channel matrix; conversely, when Type-II \gls{csi} is adopted, the feedback is composed of a linear combination of two or more (up to $4$ per polarization) vectors taken from the oversampled DFT codebook.
In this latter case, both the indices of the chosen vectors and the linear combination coefficients are fed back to gNB.
Finally, it is worth observing that the accuracy of a Type-II CSI feedback scheme is larger, and so is the resulting overhead~\cite{38214}.
The reason behind such mechanisms is to be found in the attempt to reduce the amount of \gls{csi} acquisition overhead while exploiting \gls{mimo} advantages, such as spatial multiplexing and beamforming.
Although at the moment full \gls{csi}\footnote{According to the 3GPP terminology, the term full \gls{csi} is known as explicit \gls{csi}.} is not included in the standard, ongoing discussions are attempting to assess the trade-off between precoding gain and overhead cost.

Differently from the prior art, the objective of this study is twofold.
First, we aim at evaluating the performance of diverse precoders when a realistic channel is considered, where ``realistic" denotes both the adoption of a channel model supported by experimental evidence and the inclusion of \gls{csi} imperfections.
Second, we compare the aforementioned linear precoders against \gls{gob} optimization approaches, with the goal of assessing the gain of linear precoders overs simpler (and less demanding in term of \gls{csi}) \gls{gob} approaches.

\begin{figure}[t!]
\centering
\includegraphics[width=0.8\columnwidth]{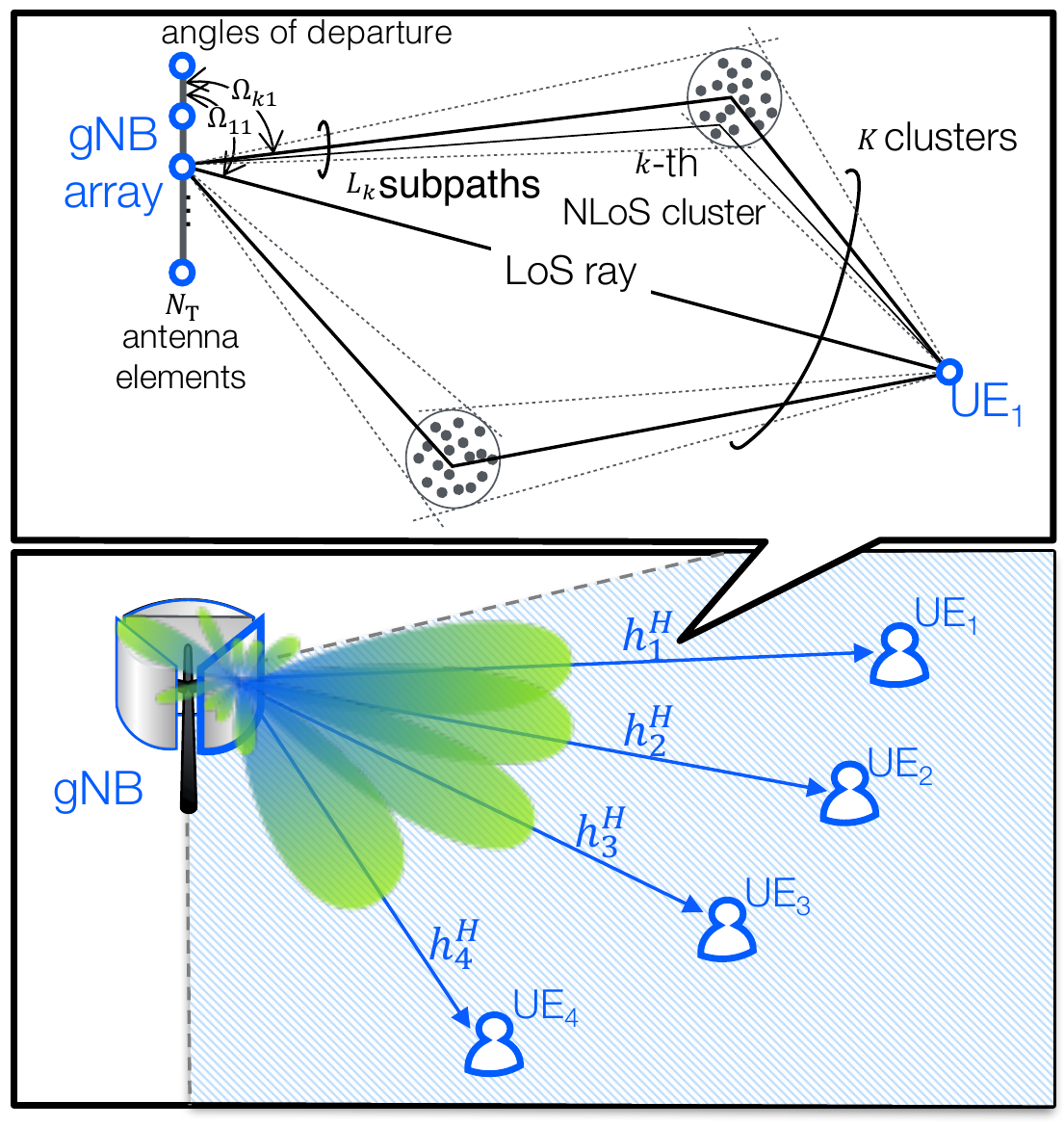}
\caption{Illustration of the \gls{mmwave} system model considered (bottom) and representation of the channel model used for each link in the framework (top).
\vspace{-0.7cm}}
\label{fig_network_overview}
\end{figure}

As reported in Figure~\ref{fig_network_overview}, we consider a scenario with both a realistic sectorization and an antenna array radiation pattern, as suggested by the 3GPP specifications in~\cite{antenna_3gpp}.
Moreover, two measurement-based realistic channel models are considered, one from \gls{nyu}~\cite{akdeniz14} and one from \gls{3gpp} as suggested in~\cite{mmWave_3gpp_channel}, both used to evaluate and compare the performance of different precoders. 

\paragraph*{Notation}
In this paper, column vectors and matrices are respectively denoted by boldface lowercase and uppercase letters. 
We identify with $\mathbf{X}^H$ the conjugate transpose of $\mathbf{X}$, and the Frobenius norm is denoted $\|\cdot\|_{\text{F}}$ while the Euclidean norm is denoted as $\|\cdot\|$.
The set of all complex numbers is denoted by $\mathbb{C}$, with $\mathbb{C}^{N \times 1}$ and $\mathbb{C}^{N \times M}$ being the generalizations to vectors and matrices, respectively.
The $M \times M$ identity matrix is written as $\mathbf{I}_M$ and the zero matrix of size $N_{\text{T}} \times M$ is denoted as $\mathbf{0}_{N_{\text{T}} \times M}$.
Finally, we generally indicate with $\widehat{\mathbf{X}}$ the Frobenius normalized matrix of $\mathbf{X}$.

\section{System Model}
\label{system_model}
We consider a narrowband single-cell downlink multi-user \gls{mimo} mmWave system where a single \gls{gnb} sector with $N_{\text{T}}$ transmit antennas is serving $M$ single-antenna \glspl{ue}.\footnote{We note that the number of \glspl{ue} that can be simultaneously supported by the \gls{gnb} sector is less than or equal to the number of antenna elements, i.e., $M \le N_\text{T}$.}
The channel to the $m$-th user is assumed narrowband and is described by the vector of coefficients $\textbf{h}_m \in \mathbb{C}^{N_\text{T} \times 1}$, and its $j$-th element describes the channel response between the $j$-th transmitting antenna element and the receive antenna.
This input-output relationship can be described as
\begin{equation}
y_{m}=\mathbf{h}_{m}^{H}\mathbf{x} + n_{m},\quad m \in \{1,2,\dots ,M\}
\end{equation}
where $\mathbf{x}$ is the $N_\text{T}\times 1$ transmitted vector signal, $y_{m} \in \mathbb{C}$ is the received signal, and $n_{m}$ is the noise term.
Assuming to use a precoder, the transmitted vector signal is $\mathbf{x} =\sum_{i=1}^{M}\mathbf{w}_{i} s_{i}$, where $s_{i}$ is the data symbol and $\mathbf{w}_{i}$ is the $N_\text{T}\times 1$ linear precoding vector.

Aggregating together the precoding vectors of all the $M$ \glspl{ue} we can define the precoding matrix $\mathbf{W} = \left[\mathbf{w}_1,\dots,\mathbf{w}_M \right] \in \mathbb{C}^{N_{\text{T}}\times M}$.
We note that, in order to respect the power constraint $\mathbb{E}\left[ \| \mathbf{W}\mathbf{s}\|^2 \right] = 1$, we normalize the precoding matrix with the Frobenius norm as follows $\widehat{\mathbf{W}} = \frac{\mathbf{W} } {\|\mathbf{W}\|_{\text{F}}}$.
Using this notation, it is possible to write the system input-output equation as
\begin{equation}
\mathbf{y} ={\mathbf{H}}^H  \widehat{\mathbf{W}} \mathbf{s} + \mathbf{n}
\end{equation}
where $\mathbf{y},\mathbf{s}$ and $\mathbf{n}$ are vectors with dimension $M\times 1$,
while channel matrix ${\mathbf{H}}$ is defined in $\mathbb{C}^{N_T\times M}$.

Finally, we define $\bar{\mathbf{H}}^{(p)}$ as the $M \times M$ equivalent matrix obtained with the product
\begin{equation}
\bar{\mathbf{H}}^{(p)} = {\mathbf{H}}^H\widehat{\mathbf{W}}^{(p)}
\label{equivalent_matrix}
\end{equation}
where superscript $p$ is used to identify the different precoding approaches evaluated as described in the following.

\subsection{Channel Models}
\label{channel_model}

In our evaluation, \gls{mimo} channel vectors $\mathbf{h}$ are generated according to three distinct statistical channel models.
The first model under analysis is a standard Rayleigh fading channel model; the second is derived from a set of extensive measurement campaigns in New York City by \gls{nyu}--Wireless~\cite{akdeniz14}; the last model considered is the one provided by the \gls{3gpp}~\cite{mmWave_3gpp_channel}, which was obtained from multiple measurement campaigns from different research groups all around the world. 
For this study, we adopt the channel model with the settings of the \gls{uma} scenario.

Both the realistic models (i.e., \gls{nyu} and \gls{3gpp}) are based on the WINNER II channel characterization~\cite{winner2}, and consider macro-level scattering paths and sub-paths.
Some minor differences are present in the settings of the models, nevertheless, a major difference is identified in the number of paths and sub-paths considered.
The \gls{nyu} characterization has higher \textit{directionality} obtained by assuming a smaller number of paths.  
To be precise, the \gls{nyu} model considered a maximum of $4$ main paths (defined as \emph{clusters}), while the \gls{3gpp} channel model can reach a maximum of $20$ clusters in \gls{nlos} conditions.
Here, by \textit{directionality}, we mean the ability of the channel and beamforming, to focus the power in a specific direction.
Together with the channel, also a measurement-based distance-dependent path loss model is considered with \gls{los}, \gls{nlos} and outage conditions.

At the transmitter side, we model the antenna of the considered sector as a \gls{upa}.
In this manner, the beamforming can be performed in both the azimuth and elevation dimensions.
Furthermore, we precisely model each element radiation pattern following the 3GPP specifications in~\cite{antenna_3gpp} and~\cite{mmWave_3gpp_channel}, as done in our previous work~\cite{rebato18}.
We consider the superposition of element radiation pattern and array factor in order to gather a precise knowledge of the array radiation effects due to beamforming.
This permits a careful characterization of the steering beams, and therefore a precise knowledge of the amount of power irradiated by the antenna array in all directions.
Thus, we are realistically computing both the desired and the interfering signals.
A complete explanation of the relationship between array and element patterns can be found in~\cite{antenna_3gpp} and~\cite{rebato18}.

The channel of each link is computed with a set of $K$ clusters and $L_k$ sub-paths per cluster (as shown in the top part of Figure~\ref{fig_network_overview}), and is represented as
\begin{align}
\mathbf{h}= \sum_{k=1}^{K}\sum_{l=1}^{L_k}g_{kl} \mathsf{F}_{\text{T}}\left(\Omega_{kl}\right) \textbf{u}_{\text{T}}\left(\Omega_{kl}\right)
\label{channel_vector}
\end{align}
where {$g_{kl}$} is the small-scale fading gain of sub-path $l$ in cluster $k$, and $\textbf{u}_{\text{TX}}$ is the 3D spatial signature vector of the transmitter.
Furthermore, for brevity, we use subscript or superscript T, referring to a transmitter related term.
Moreover, $\Omega_{kl} = \left( \theta_{kl}, \phi_{kl}\right)$ is the angular spread of vertical and horizontal \gls{aod} for sub-path $l$ in cluster $k$~\cite{akdeniz14}.
Finally, $\mathsf{F}_{\text{T}}$ is the field factor term of the transmitting array.
Detailed explanation on how to precisely compute all these channel terms can be found in~\cite{akdeniz14} and~\cite{rebato18}. 

\vspace{-0.5cm}
\subsection{Precoders Considered} 
\label{precoders}
With the intent to perform a study of the different precoding techniques while realistically modeling the channel, we discuss in the following paragraphs all the approaches evaluated and provide details on how they are computed.

\paragraph{Grid of beams (power optimization)}
This approach consists in the use of a codebook $\mathcal{Z}$ of precomputed precoders that will be tested with the aim to choose the one that maximizes a specific metric.
Each precoder vector in the codebook represents a \gls{dft} beam pointing towards a direction.
According to this principle, the entire codebook spans the whole effective area.\footnote{This principle is an assumption adopted for this evaluation. Different codebook designs can also be applied in our optimization.} 

Two different \gls{gob} metrics and thus optimization criteria are considered in this study.
First, for each active \gls{ue}, we identify the precoder $\mathbf{w}_m^{(\mathsf{GoB}_{\mathsf{P}})}$ which maximizes the received power among all possible precoder vectors $z$ in the codebook $\mathcal{Z}$, thus 
\begin{equation}
\mathbf{w}_m^{(\mathsf{GoB}_{\mathsf{P}})} = \arg\max_{z \in \mathcal{Z}} | \mathbf{h}^H_m \mathbf{w}_z|^2.
\end{equation}
We identify it with the acronym $\mathsf{GoB}_{\mathsf{P}}$, and the respective precoding matrix is derived as
\begin{equation}
\mathbf{W}^{(\textsf{GoB}_{\mathsf{P}})} = \left[ \mathbf{w}_1^{(\mathsf{GoB}_{\mathsf{P}})}, \dots, \mathbf{w}_M^{(\mathsf{GoB}_{\mathsf{P}})} \right].
\label{precoder_vector_gob}
\end{equation}

\paragraph{Grid of beams (SLNR optimization)}
Similarly, we study an alternative in which the precoder is chosen by maximizing the \gls{slnr} for each single \gls{ue} $m$.
We define it as $\mathsf{GoB}_{\mathsf{SLNR}}$ and the optimization expression becomes
\begin{equation}
\mathbf{w}_m^{(\mathsf{GoB}_{\mathsf{SLNR}})} = \arg\max_{z \in \mathcal{Z}} \left ( \frac{| \mathbf{h}^H_{m,m} \mathbf{w}_z|^2}{\sigma^2 + \sum_{ i \neq m} | \mathbf{h}^H_{m,i} \mathbf{w}_z|^2}  \right)
\end{equation}
then, the precoder matrix $\mathbf{W}^{(\textsf{GoB}_{\mathsf{SLNR}})}$ is derived as in~\eqref{precoder_vector_gob}.

The rationale behind this choice is that the sum of \glspl{slnr} is a close approximation of the sum of \glspl{sinr}, with the advantage of being computationally much easier to perform.
This stems mainly from the fact that whereas the sum \gls{sinr} maximization would require an exhaustive search for all possible beams and all users in the cell, the sum \gls{slnr} can be maximized by simply maximizing the \gls{slnr} of each \gls{ue}.
	

\paragraph{Matched filter precoder~\cite{joham05}}
The \gls{mf}, also known as conjugate beamforming, maximizes the power of the received signal, without any interference consideration.
It is optimum when the noise power received by the \gls{ue} is much stronger than the interference that would result from the transmitted signals intended to be received by the co-scheduled \glspl{ue}.
For this reason, it is optimum for noise-limited scenarios.\footnote{\emph{Noise-limited} and \emph{interference-limited} scenario refer, respectively, to the case in which the noise power is greater than the interference power and vice-versa.}
Its precoding matrix is expressed as
\begin{equation}
\mathbf{W}^{(\textsf{MF})} = \widehat{\mathbf{H}}.
\end{equation}
The \gls{gnb} computes the precoding matrix after estimating the channel so as to direct the useful energy in the direction of each \gls{ue}.
In our evaluation, we assume complete knowledge of the channel and we use this assumption for the calculation of this and the next precoders.
\begin{figure*}[t!]
    \centering
    \begin{subfigure}[b]{0.3\textwidth}
        \includegraphics [width=.9\textwidth,clip]{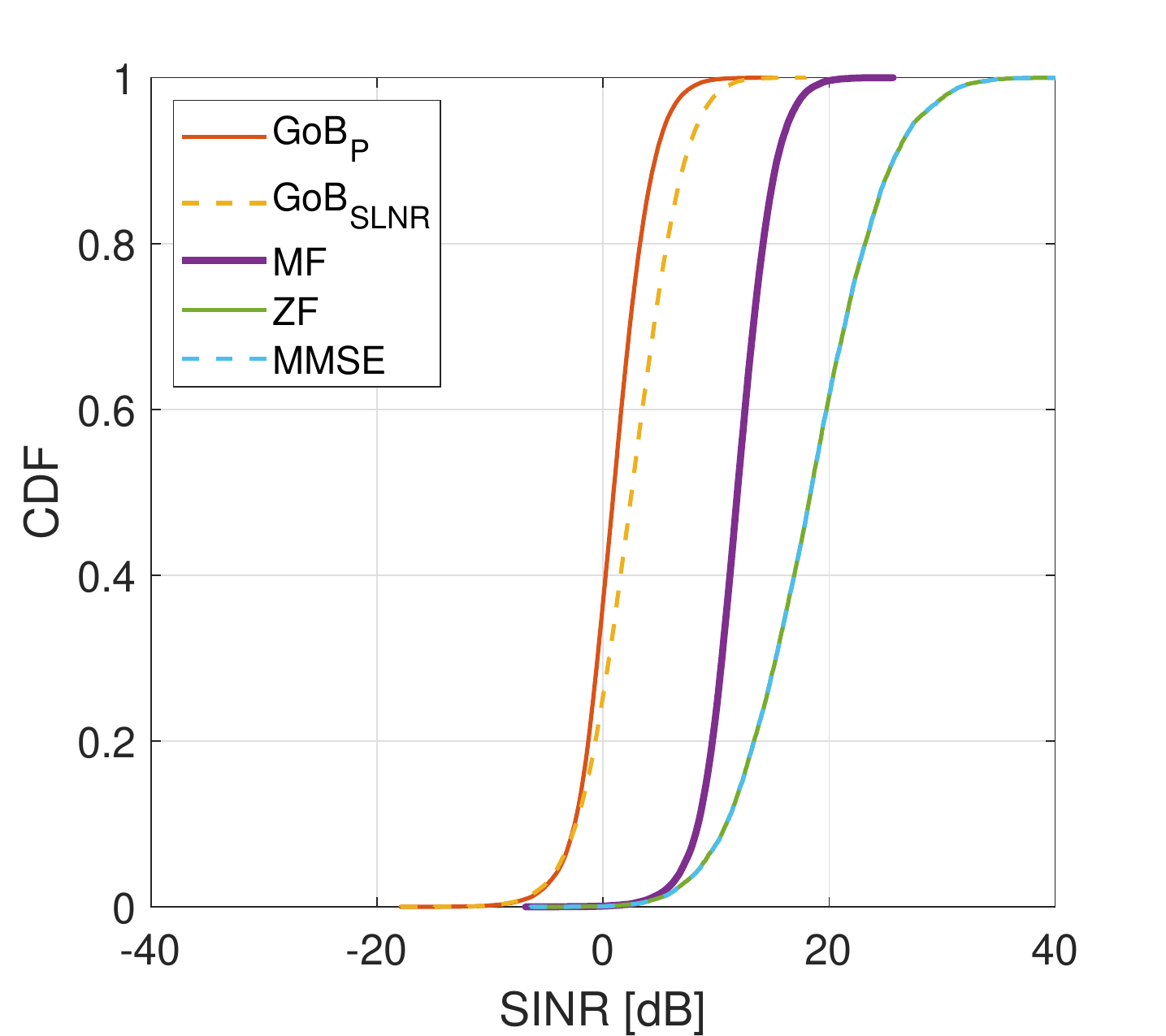}
        \caption{Rayleigh fading channel.}
        \label{fig_cdf_random}
    \end{subfigure}
    ~ 
    \begin{subfigure}[b]{0.3\textwidth}
        \includegraphics[width=.9\textwidth,clip]{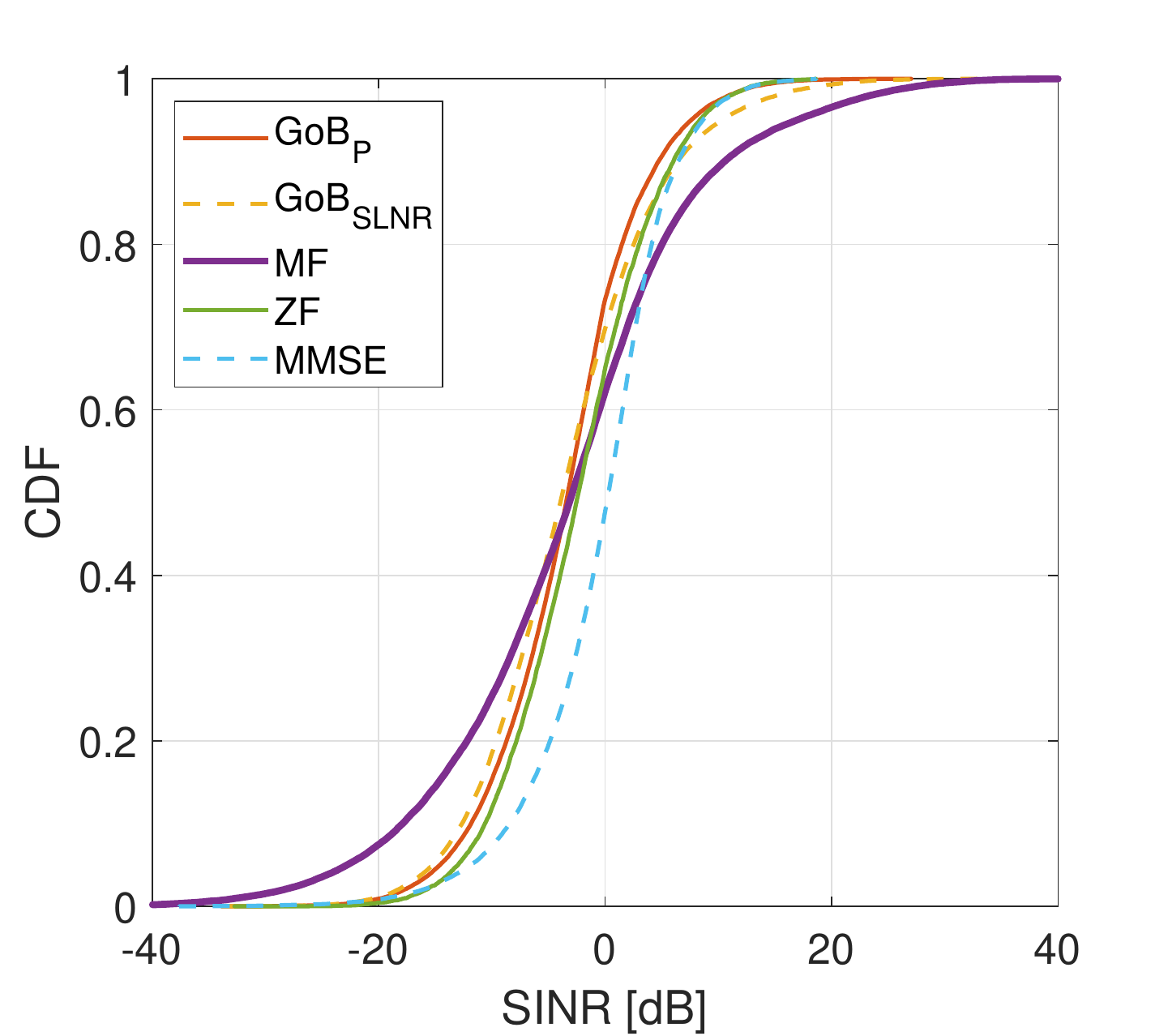}
        \caption{\gls{nyu} channel model.}
        \label{fig_cdf_nyu}
    \end{subfigure}
    ~ 
    \begin{subfigure}[b]{0.3\textwidth}
        \includegraphics[width=.9\textwidth,clip]{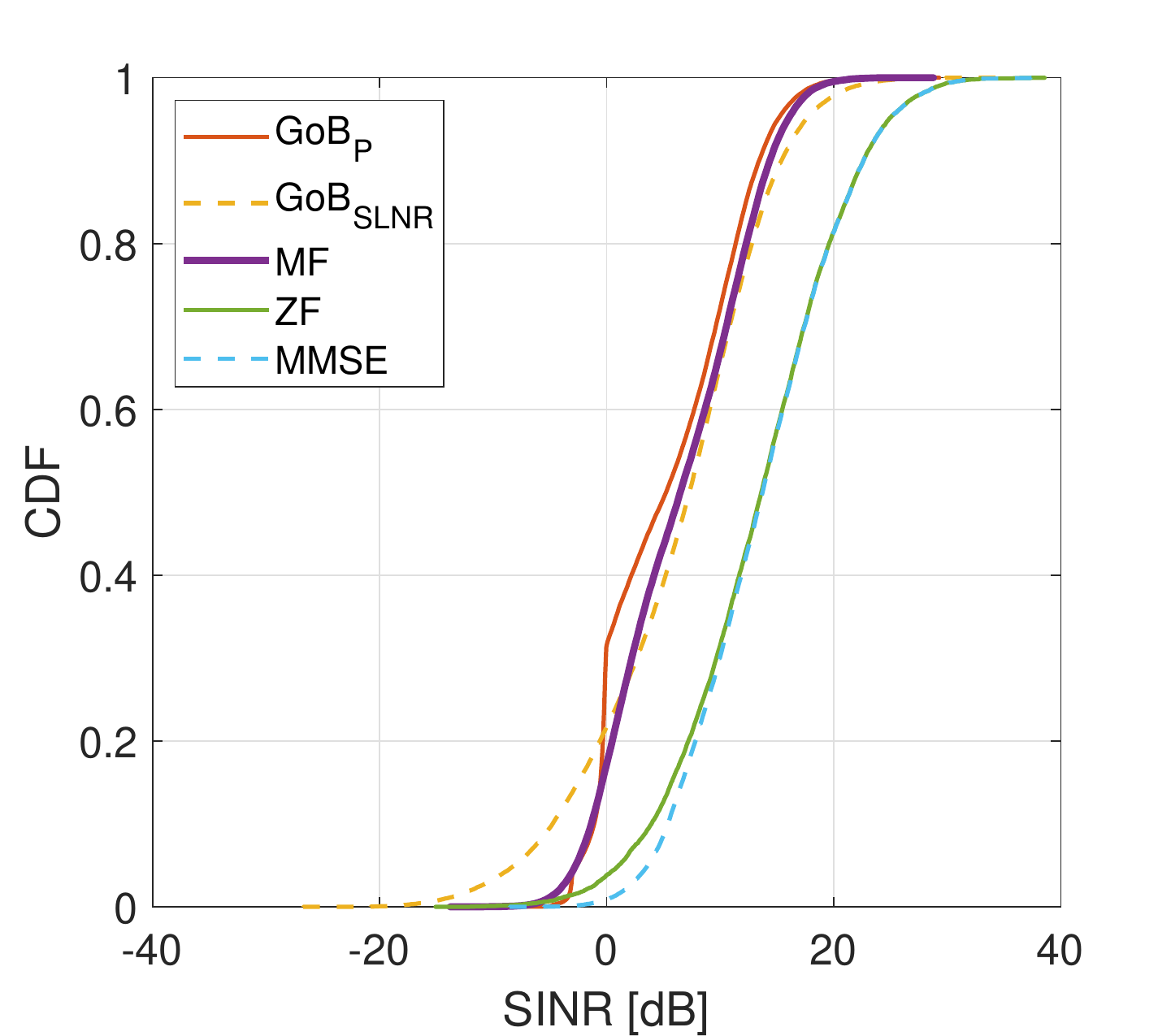}
        \caption{\gls{3gpp} channel model.}
        \label{fig_cdf_3gpp}
    \end{subfigure}
    \glsunset{cdf}
    \caption{Empirical \gls{cdf} of the \gls{sinr} for the different precoders evaluated. In these figures, we used a fixed number of \glspl{ue} $M=4$ and $P_{\text{T}} = 30$~dBm.
    \vspace{-0.3cm}}
    \glsreset{cdf}
    \label{fig:global_cdf}
\end{figure*}

\paragraph{Zero-forcing precoder}
An evolution of \gls{mf} linear processing can be used to limit the detrimental effects of multi-user interference.
The \gls{zf} precoder tries to cancel the power of the interference, and therefore is an optimal solution for interference-limited scenarios.
This interference canceling property is obtained at the price of a slightly complex precoder computation and of a reduced received power.
The precoding matrix is designed according to the \gls{zf} criterion~\cite{zero-forcing} and is given by
\begin{equation}
\mathbf{W}^{(\textsf{ZF})} = \widehat{\mathbf{H}} (\widehat{\mathbf{H}}^H\widehat{\mathbf{H}})^{-1}
\end{equation}
which simply denotes the right pseudo-inverse of the matrix $ \mathbf{\widehat{H}}^H$.
\paragraph{MMSE precoder}
Differently from the last two precoders considered, the \gls{mmse} precoding strategy (also known as Kalman filter precoder) maximizes the sum of the $\textrm{SINR}$.
Therefore, it optimizes the received power while minimizing the interference signal. 
It can be considered as a solution in between the \gls{mf} and the \gls{zf} precoders.
The precoding matrix is expressed as
\begin{equation}
\mathbf{W}^{(\textsf{MMSE})} = \widehat{\mathbf{H}} \left(\widehat{\mathbf{H}}^H\widehat{\mathbf{H}} + \frac{1}{\textrm{SNR}} \mathbf{I}_M \right)^{-1}
\end{equation}
and it is possible to prove that it can be expressed as a linear combination of \gls{mf} and \gls{zf} precoders~\cite{Rose16}.  

\subsection{Imperfect channel estimate}
\label{imperfect_channel}

Focusing on a realistic system, achieving a complete and correct knowledge of the \gls{csi} is not feasible in a practical framework.
To be precise, typical mmWave implementation does not have direct access to the signals received on each \gls{gnb} antenna, so learning the channel on each antenna element is currently extremely difficult and almost infeasible.
For this reason, we  consider the performance in case the transmitter has an imperfect channel estimate.
The channel estimation error is modeled following a Gauss-Markov formulation, where the imperfect channel $\mathbf{H}_e$ is obtained using the \emph{true} channel $\mathbf{H}$ as follows
\begin{equation}
\mathbf{H}_e = \tau \mathbf{H} + \sqrt{1-\tau^2} \mathbf{E} 
\end{equation}
where each term of the matrix $\mathbf{E}$ follows a circularly symmetric Normal distribution $\mathcal{CN}(0,1)$.
Moreover, the scalar parameter $\tau \in [0,1]$ is used to indicate the quality of the channel estimation, where $\tau = 1$ corresponds to perfect estimation of the channel whereas $\tau = 0$ corresponds to having only the random channel $\mathbf{E}$~\cite{muller14}. 
This parameter depends on factors such as the time/power spent on pilot-based channel estimation.
As done in~\eqref{equivalent_matrix}, and with the imperfect channel consideration, the equivalent matrix becomes $\bar{\mathbf{H}}_e = \widehat{\mathbf{H}}^H\widehat{\mathbf{W}}^{(p)}_e$, where the precoder $\widehat{\mathbf{W}}^{(p)}_e$ has been calculated considering the imperfect channel $\mathbf{H}_e$.

\section{Comparison results}
\label{comparison_results}

\begin{table}
\centering
\caption{List of parameters used in our evaluation. Unless specified otherwise, these settings are considered as default in all the studies carried out during our evaluation.}
\small
\renewcommand{\arraystretch}{0.95}
{\begin{tabular}{r l}
\toprule
\hspace{-10pt}\bf{Value} &\hspace{-5pt} \textbf{Meaning and (Notation)}\\
\cmidrule(r){1-1}\cmidrule(r){2-2}
\hspace{-10pt} $28$~GHz & carrier frequency ($f$) \\
\hspace{-10pt} $7$~dB & noise figure ($\textsf{NF})$\\
\hspace{-10pt} $100$~m & transmitter receiver distance \\
\hspace{-10pt} $4$ & \# of served \glspl{ue} ($M$)\\
\hspace{-10pt} $64$ & total \# of antennas per \gls{gnb} ($N_\text{T})$\\
\hspace{-10pt} $[8 \times 8]$ & vertical and horizontal \gls{upa} configuration\\
\hspace{-10pt} $\lambda[0.7,0.5] $ & vertical and horizontal \gls{upa} element spacing \\
\hspace{-10pt} $30$~dBm & transmitted power ($P_{\text{T}}$)\\
\hspace{-10pt} $6$ & \# bits phase shifters \\
\hspace{-10pt} $0.99$ & imperfect channel metric ($\tau$)\footnotemark \\
\bottomrule
\end{tabular}}
\label{table_notations}
\vspace{-0.5cm}
\end{table}
\footnotetext{The value $\tau =0.99$ identifies an optimistic channel imperfection.
As discussed later in the results, even with a small error in the \gls{csi} the degradation is notable.}
In this section we provide some simulation results to compare the performance of the different precoders considered, which will be assessed in terms of \gls{sinr} and achievable system capacity.
Before examining in detail all the results, we briefly report here the \gls{sinr} expression used in our evaluation.
Furthermore, Table~\ref{table_notations} details all the parameters and respective values adopted.

The first metric considered in our evaluation is the \gls{sinr}, we calculate it for each \gls{ue} $m$ as follows
\begin{equation}
\textrm {SINR}_{m}^{(p)} = \frac{|\bar{\mathbf{h}}^{(p)}_{m,m}|^2}{ \tfrac{1}{\textrm{SNR}} + \sum_{ i \neq m } |\bar{\mathbf{h}}^{(p)}_{m,i}|^2 }
\label{equation_sinr}
\end{equation}
where $\textrm{SNR}$ is computed using the transmitted power, the path loss $\ell$, and the thermal noise $\sigma^2$ as $\tfrac{P_{\text{T}} \ell^{-1}}{\sigma^2}$.
We note that each UE's \gls{sinr} is affected by the accurate antenna array radiation pattern that is computed considering the field factor term into the channel gains, as previously described in~\eqref{channel_vector}.
Finally, superscript $(p)$ is used to identify the $M \times 1$ vector of the equivalent matrix $\bar{\mathbf{H}}^{(p)}$ obtained with the corresponding precoder $\widehat{\mathbf{W}}^{(p)}$.
We recall that the precoding matrix is included into the equivalent matrix as done in~\eqref{equivalent_matrix}.

The first result is a comparison of the \gls{sinr} values for all the precoders studied under different channel model assumptions.
In order to have a comprehensive view of the overall performance, Figure~\ref{fig:global_cdf} reports the empirical \gls{cdf} of the \gls{sinr} for all the configurations considered.
The results have been collected over a sufficient number of repetitions in order to obtain the desired accuracy, thus precisely evaluating the different precoders.
We compare both the \gls{nyu} and \gls{3gpp} channel characterizations with a random Rayleigh channel model computed as $\mathbf{H}_{\text{R}} \sim \tfrac{1}{\sqrt{M}}\mathcal{CN}\left( \boldsymbol{0}_{N_{\text{T}} \times M},\boldsymbol{1}_{N_{\text{T}} \times M} \right)$.
The figures are obtained with a Monte Carlo approach which generates random samples of channel and environment for all the \glspl{ue} in each iteration.
As expected, the \gls{mmse} precoder outperforms all the other configurations for most of the \glspl{ue}. 
Furthermore, due to the high directivity of the \gls{nyu} channel, both \gls{gob} precoders are able to reach higher \gls{sinr} values, with respect to the \gls{mf}, for more than forty percent of the \glspl{ue}.
We note that for this plot we have used a $7$~dB noise figure\footnote{The noise figure term quantifies the degradation of the \gls{snr} due to the noise present in the system.}, which corresponds to a mostly interference-limited system~\cite{rebato16}. 
Due to lack of space, we are not reporting here any results with higher noise power.
However, larger noise values push the system into a noise-limited regime, hence reducing the performance of the \gls{zf} precoder.

\begin{figure}[t!]
\centering
\includegraphics[width=.8\columnwidth]{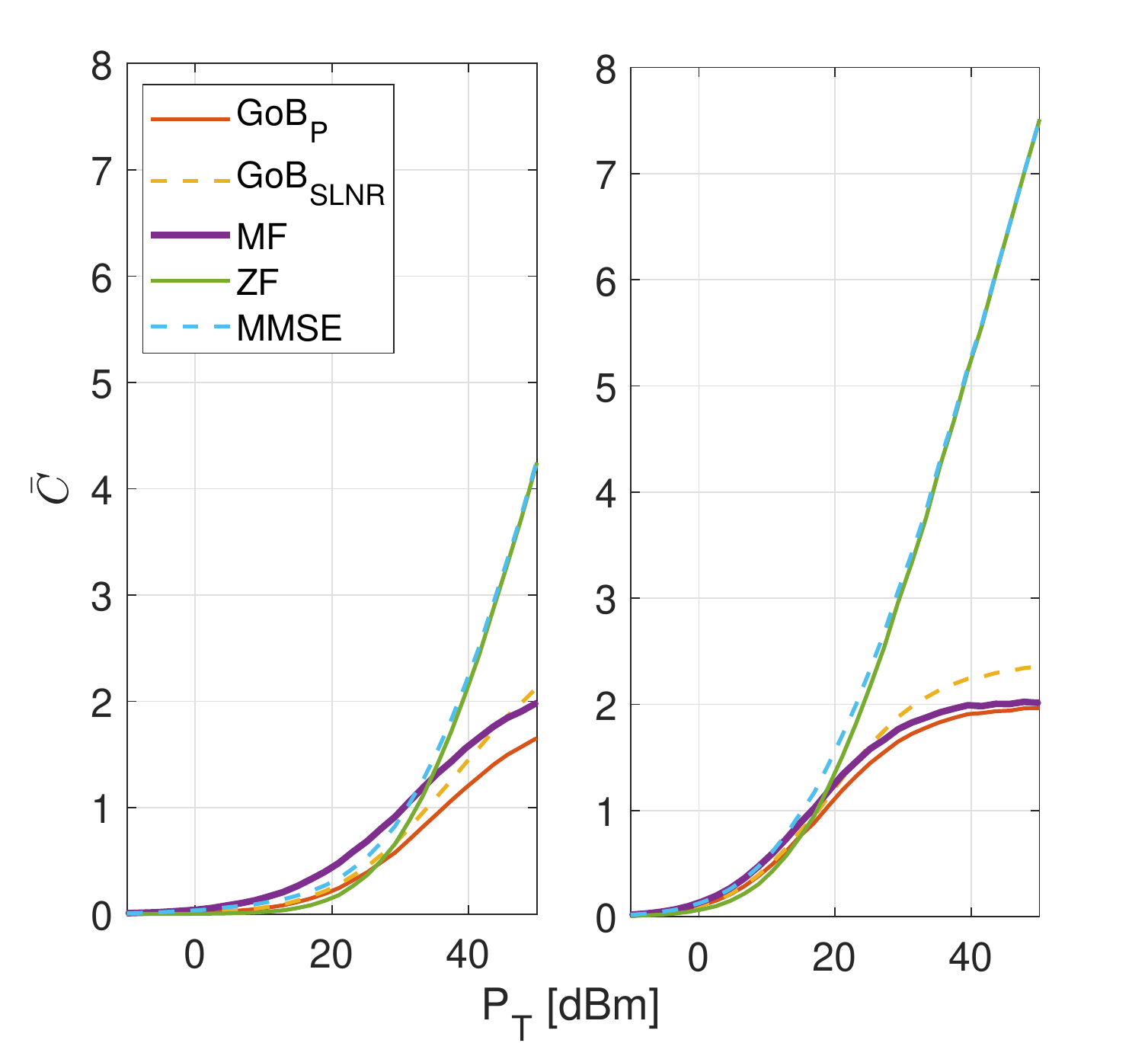}
\caption{System capacity for the different precoders, varying the transmitted power $P_{\text{T}}$. In this figure, $M=4$, $\textsf{NF} = 7$~dB, and the channel is modeled as~\gls{nyu} in the left plot and following the \gls{3gpp} characterization in the right plot.
\vspace{-0.5cm}}
\label{fig_capacity}
\end{figure}
Figure~\ref{fig:global_cdf} also highlights the lack of fairness among \glspl{ue} when the \gls{mf} precoder is used.
Even if the average value of \gls{mf}'s \gls{sinr} is the highest, really high values of \gls{sinr} are obtained only for a small percentage of \glspl{ue}.  
As we can see in Figure~\ref{fig_cdf_random}, the precoders which do not require knowledge of the channel (e.g., $\mathsf{GoB}_{\mathsf{P}}$ and $\mathsf{GoB}_{\mathsf{SLNR}}$) are unable to operate efficiently when the channel is Rayleigh.
Conversely, they show \gls{sinr} values close to those of \gls{mf} when the channel is modeled following the \gls{nyu} characterization.
This stems from the fact that, while a Rayleigh channel is \emph{isotropic} and scatters the power without a preferred direction, realistic mmWave channels present limited multi-paths and rays with a large portion of the power concentrated in few directions.
To further support such interpretation, Figure~\ref{fig_cdf_nyu} reports the \gls{sinr} values under \gls{nyu} and \gls{3gpp} \gls{uma} channel characterizations.
We note that, since the \gls{nyu} model has fewer clusters than its \gls{uma} counterpart, the power is concentrated in fewer directions and thus the \gls{gob} approaches have even higher performance.

With the use of the \gls{sinr} expression in~\eqref{equation_sinr}, we can compute the channel capacity as follows
\begin{equation}
\textrm{C}^{(p)}_m = \log _{2} \left(1+\textrm {SINR}^{(p)}_{m} \right).
\label{equation_capacity}
\end{equation}
This metric can be used to evaluate the spectral efficiency of each configuration, and we  indicate its average value by $\bar{\textrm{C}}$.
Figure~\ref{fig_capacity} plots the average system capacity in the different configurations as a function of the transmit power $P_{\text{T}}$ used at the \gls{gnb} side.
The growing transmit power increases at the same rate the received power and the interference levels for the interference-blind precoders (i.e., \gls{mf}, \gls{gob}) hence resulting in a saturation of the performance. 
Conversely, interference-aware precoders such as \gls{zf} and \gls{mmse} can have indefinitely growing performance.
	%
The figures display the good performance of the \gls{gob} precoder when the \gls{slnr} is optimized. 
We remark that, even if the average \gls{mf} \gls{sinr} is higher with respect to the other configurations, it presents poor fairness as previously discussed.

Comparing the two realistic models, we can notice also in these figures how the directivity of the \gls{nyu} model results in good outcomes for the capacity in the two \gls{gob} approaches in the range of values around $20$~dBm of transmitted power.
Contrary to expectation, in this particular range, \gls{gob} procedures can perform appreciably better than \gls{zf} if high spectral efficiency is desired, while, if a more energy-efficient operating point is chosen, the performance gap narrows, and eventually \gls{mf} outperforms all the other configurations.

\begin{figure}[t!]
\centering
\includegraphics[width=.8\columnwidth,clip]{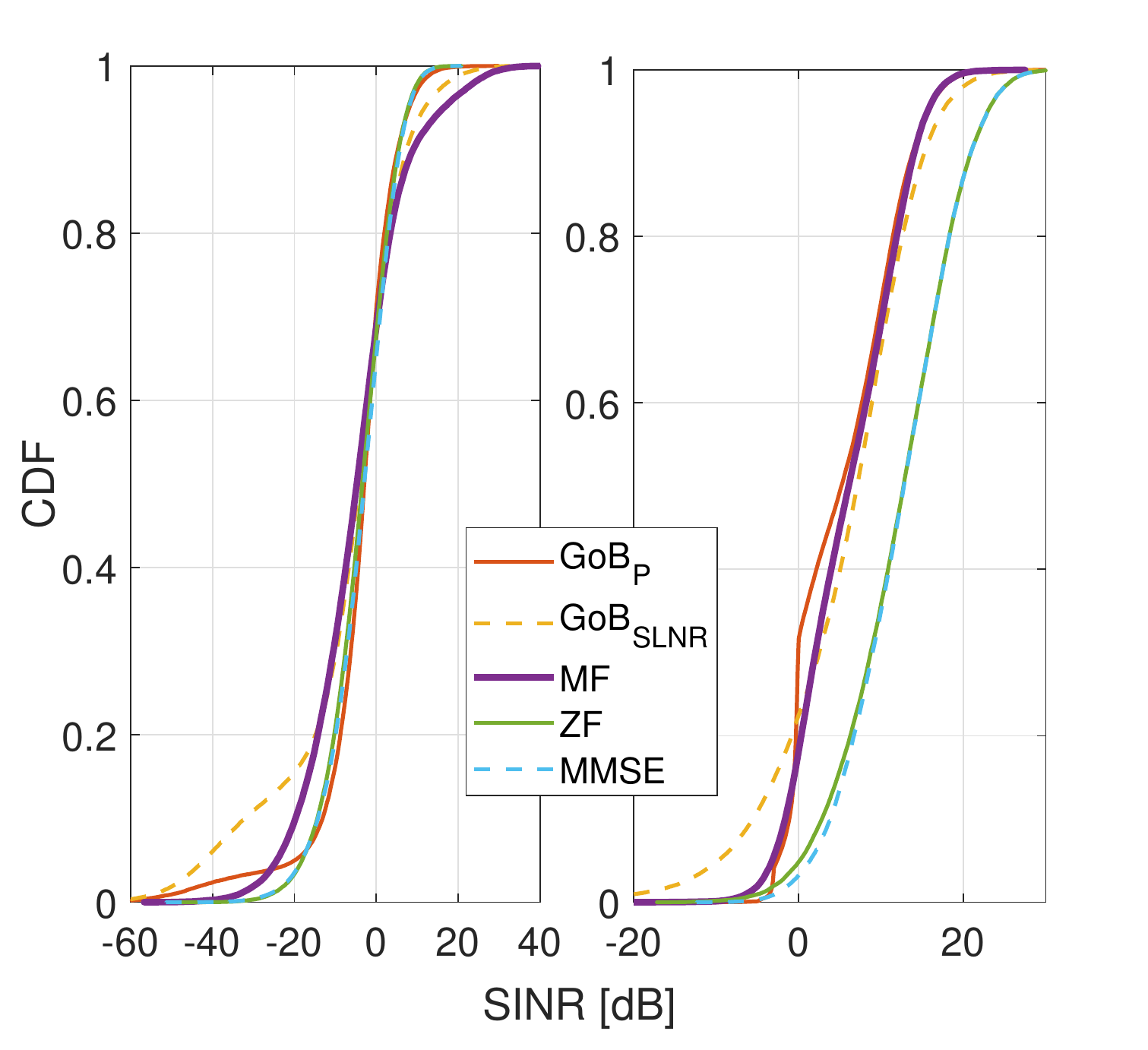}
\caption{Empirical \gls{cdf} of the \gls{sinr} for the different precoders evaluated when an imperfect channel is considered. In this figure, $M=4$, $\textsf{NF} = 7$~dB and $\tau = 0.99$, and the channel is modeled as \gls{nyu} in the left plot and following the \gls{3gpp} characterization in the right plot.
\vspace{-0.5cm}}
\label{fig_error}
\end{figure}

As the last result, Figure~\ref{fig_error} reports the empirical \gls{cdf} of the \gls{sinr} when an imperfect channel $\mathbf{H}_e$ is considered.
Comparing it with Figure~\ref{fig:global_cdf}, is it possible to notice how the imperfection in the \gls{csi} results in a degradation of the performance, especially for the linear precoders.
More importantly, when an error in the channel estimation is considered, the gap between \gls{gob} optimization approaches and linear precoders is strongly reduced and, in most cases, \gls{gob} is even able to outperform the linear precoders.
We recall that gathering the \gls{csi} necessary to use \gls{mf}, \gls{zf} and the \gls{mmse} precoders has a cost for the system that should be properly considered.
Furthermore, given the small implementation loss of \gls{gob} precoding with respect to more refined systems, and considering the high level of complexity that gathering the necessary \gls{csi} would require, it seems that the additional complexity may not be justified by the modest (or even vanishing) performance improvement.

\subsubsection*{Remarks}
\label{results_remarks}
We report in this subsection the main remarks raised in our evaluation study. 
Due to the directionality of \gls{mmwave} channels, our results support \gls{gob} approaches as a good trade-off between \gls{csi} acquisition complexity and performance. 
Given the limited advantage (about $+4$~dB with \gls{mmse} and the \gls{nyu} channel model), there is no strong motivation to use linear precoders in multi-user systems at \gls{mmwave} frequencies.
Table~\ref{table_results} summarizes our findings, reporting the \gls{sinr} gaps for the $50^{\text{th}}$ percentile in the different approaches considered.
Although linear precoders can exploit a larger amount of information on the channel matrix, requiring full \gls{csi} at the transmitter, the gain under imperfect \gls{csi} can be assessed as less than $+1.36$~dB with \gls{mmse} and the \gls{nyu} channel model.
If the inaccuracy of the channel estimation grows, it is possible to conjecture that the gap would close even more, eventually eliding any advantage.
A similar trend, though with slightly higher gains, can be observed when the \gls{3gpp} channel model is considered.
\begin{table}[]
\centering
\vspace{0.2cm}
\caption{Evaluation of the gaps in the $50^{\text{th}}$ percentile of the \gls{sinr} expressed in dB for the different precoders considered in this evaluation.
Table obtained with a fixed number of \glspl{ue} $M=4$, and $P_{\text{T}} = 30$~dBm.}
\label{table_results}
\small
\renewcommand{\arraystretch}{0.95}
\begin{tabular}{rcccc}
\toprule
\multicolumn{1}{l}{} & \multicolumn{2}{c}{\begin{tabular}[c]{@{}c@{}}perfect \gls{csi}\\ ($\tau = 1$)\end{tabular}} & \multicolumn{2}{c}{\begin{tabular}[c]{@{}c@{}}imperfect \gls{csi}\\ ($\tau = 0.99$)\end{tabular}} \\
\midrule
\multicolumn{1}{l}{} & \gls{nyu}      & \gls{3gpp}         & \gls{nyu}  & \gls{3gpp}    \\
\midrule
$\textsf{MF}$ -- $\textsf{GoB}_{\mathsf{SLNR}}$    & $+ 0.87$  & $- 0.80$    & $- 0.46$  & $- 1.03$  \\
$\textsf{ZF}$ -- $\textsf{GoB}_{\mathsf{SLNR}}$    & $+ 1.40$  & $+ 6.33$    & $+ 0.89$  & $+ 5.67$  \\
$\textsf{MMSE}$ -- $\textsf{GoB}_{\mathsf{SLNR}}$  & $+ 4.05$  & $+ 6.45$    & $+ 1.36$  & $+ 5.67$  \\                         
\bottomrule
\end{tabular}
\vspace{-0.3cm}
\end{table}

\section{Conclusion and future works}
\label{conclusion_and_future_works}
In this study, we have highlighted the impact of realistic mmWave channel behaviors on the downlink \gls{mmwave} \gls{mu} \gls{mimo} system when different precoders are considered at the \gls{gnb} side.
Our study led to the following observations.
Under ideal condition (i.e., Rayleigh channel model, perfect \gls{csi}), linear precoders largely outperform \gls{gob} due to their ability to perfectly adapt to the channel realization.
However, the directionality present in realistic channel models reduces the gap, sometimes even letting the \gls{gob} approaches surpass more complex solutions and in most cases not justifying the additional complexity.

Finally, we have studied how the performance behaves when considering an error in the \gls{csi} acquisition.
Results show that, even with a small \gls{csi} imprecision, the performance gap between linear precoding and \gls{gob} vanishes.

The study of more refined \gls{mmse} approaches which include per-user power balancing, as well as a theoretical evaluation of the performance of the various techniques, also for different frequency bands, are left for future study.

\bibliographystyle{IEEEtran}
\bibliography{biblio}

\end{document}

%% file: acronyms.tex
\newacronym{uma}{UMa}{Urban Macro}
\newacronym{mu}{MU}{Multi-User}
\newacronym{dft}{DFT}{Discrete Fourier Transform}
\newacronym{gob}{GoB}{Grid of Beams}
\newacronym{nyu}{NYU}{New York University}
\newacronym{mmse}{MMSE}{Minimum Mean Square Error}
\newacronym{slnr}{SLNR}{Signal to Leakage plus Noise Ratio}
\newacronym{aod}{AoD}{Angles of Departure}
\newacronym{mf}{MF}{Matched Filtering}
\newacronym{zf}{ZF}{Zero Forcing}
\newacronym{upa}{UPA}{Uniform Planar Array}
\newacronym{qos}{QoS}{Quality of Service}
\newacronym{lsa}{LSA}{Licensed Spectrum Access}
\newacronym{3gpp}{3GPP}{3rd Generation Partnership Project}
\newacronym{adc}{ADC}{Analog to Digital Converter}
\newacronym{5g}{5G}{5th generation}
\newacronym{aimd}{AIMD}{Additive Increase Multiplicative Decrease}
\newacronym{am}{AM}{Acknowledged Mode}
\newacronym{amc}{AMC}{Adaptive Modulation and Coding}
\newacronym{aqm}{AQM}{Active Queue Management}
\newacronym{awgn}{AGWN}{Additive White Gaussian Noise}
\newacronym{balia}{BALIA}{Balanced Link Adaptation}
\newacronym{bdp}{BDP}{Bandwidth-Delay Product}
\newacronym{bf}{BF}{Beamforming}
\newacronym{cc}{CC}{Congestion Control}
\newacronym{cdf}{CDF}{Cumulative Distribution Function}
\newacronym{cn}{CN}{Core Network}
\newacronym{cqi}{CQI}{Channel Quality Information}
\newacronym{cp}{CP}{Control Plane}
\newacronym{csi}{CSI}{Channel State Information}
\newacronym{csirs}{CSI-RS}{Channel State Information - Reference Signal}
\newacronym{dc}{DC}{Dual Connectivity}
\newacronym{dce}{DCE}{Direct Code Execution}
\newacronym{dci}{DCI}{Downlink Control Information}
\newacronym{dl}{DL}{Downlink}
\newacronym{dmr}{DMR}{Deadline Miss Ratio}
\newacronym{dmrs}{DMRS}{DeModulation Reference Signal}
\newacronym{e2e}{E2E}{End-to-End}
\newacronym{ecn}{ECN}{Explicit Congestion Notification}
\newacronym{edf}{EDF}{Earliest Deadline First}
\newacronym{enb}{eNB}{evolved Node Base}
\newacronym{epc}{EPC}{Evolved Packet Core}
\newacronym{es}{ES}{Edge Server}
\newacronym{fdma}{FDMA}{Frequency Division Multiple Access}
\newacronym{fdd}{FDD}{Frequency Division Duplexing}
\newacronym[firstplural=Radio Access Technologies (RATs)]{rat}{RAT}{Radio Access Technology}
\newacronym{fs}{FS}{Fast Switching}
\newacronym{ftp}{FTP}{File Transfer Protocol}
\newacronym{gnb}{gNB}{Next Generation Node Base}
\newacronym{harq}{HARQ}{Hybrid Automatic Repeat reQuest}
\newacronym{hetnet}{HetNet}{Heterogeneous Network}
\newacronym{hh}{HH}{Hard Handover}
\newacronym{hol}{HOL}{Head-of-Line}
\newacronym{ia}{IA}{Initial Access}
\newacronym{imt}{IMT}{International Mobile Telecommunication}
\newacronym{iot}{IoT}{Internet of Things}
\newacronym{los}{LoS}{Line of Sight}
\newacronym{lte}{LTE}{Long Term Evolution}
\newacronym{m2m}{M2M}{Machine to Machine}
\newacronym{mac}{MAC}{Medium Access Control}
\newacronym{mc}{MC}{Multi-Connectivity}
\newacronym{mcs}{MCS}{Modulation and Coding Scheme}
\newacronym{mec}{MEC}{Mobile Edge Cloud}
\newacronym{mi}{MI}{Mutual Information}
\newacronym{mimo}{MIMO}{Multiple-Input Multiple-Output}
\newacronym{mmwave}{mmWave}{millimeter wave}
\newacronym{mptcp}{MPTCP}{Multipath TCP}
\newacronym{mr}{MR}{Maximum Rate}
\newacronym{mss}{MSS}{Maximum Segment Size}
\newacronym{mtd}{MTD}{Machine-Type Device}
\newacronym{mtu}{MTU}{Maximum Transmission Unit}
\newacronym{nfv}{NFV}{Network Function Virtualization}
\newacronym{nlos}{NLoS}{Non Line of Sight}
\newacronym{nr}{NR}{New Radio}
\newacronym{ofdm}{OFDM}{Orthogonal Frequency Division Multiplexing}
\newacronym{pdcch}{PDCCH}{Physical Downlonk Control Channel}
\newacronym{pdcp}{PDCP}{Packet Data Convergence Protocol}
\newacronym{pdsch}{PDSCH}{Physical Downlink Shared Channel}
\newacronym{pdu}{PDU}{Packet Data Unit}
\newacronym{pf}{PF}{Proportional Fair}
\newacronym{pgw}{PGW}{Packet Gateway}
\newacronym{phy}{PHY}{Physical}
\newacronym{pbch}{PBCH}{Physical Broadcast Channel}
\newacronym[plural=\gls{mme}s,firstplural=Mobility Management Entities (MMEs)]{mme}{MME}{Mobility Management Entity}
\newacronym{prb}{PRB}{Physical Resource Block}
\newacronym{pss}{PSS}{Primary Synchronization Signal}
\newacronym{pucch}{PUCCH}{Physical Uplink Control Channel}
\newacronym{pusch}{PUSCH}{Physical Uplink Shared Channel}
\newacronym{rach}{RACH}{Random Access Channel}
\newacronym{ran}{RAN}{Radio Access Network}
\newacronym{red}{RED}{Random Early Detection}
\newacronym{rf}{RF}{Radio Frequency}
\newacronym{rlc}{RLC}{Radio Link Control}
\newacronym{rlf}{RLF}{Radio Link Failure}
\newacronym{rrc}{RRC}{Radio Resource Control}
\newacronym{rrm}{RRM}{Radio Resource Management}
\newacronym{rr}{RR}{Round Robin}
\newacronym{rs}{RS}{Remote Server}
\newacronym{rsrp}{RSRP}{Reference Signal Received Power}
\newacronym{rss}{RSS}{Received Signal Strength}
\newacronym{rtt}{RTT}{Round Trip Time}
\newacronym{rw}{RW}{Receive Window}
\newacronym{rx}{RX}{Receiver}
\newacronym{sa}{SA}{standalone}
\newacronym{sack}{SACK}{Selective Acknowledgment}
\newacronym{sap}{SAP}{Service Access Point}
\newacronym{sch}{SCH}{Secondary Cell Handover}
\newacronym{scoot}{SCOOT}{Split Cycle Offset Optimization Technique}
\newacronym{sdma}{SDMA}{Spatial Division Multiple Access}
\newacronym{sinr}{SINR}{Signal to Interference plus Noise Ratio}
\newacronym{sm}{SM}{Saturation Mode}
\newacronym{snr}{SNR}{Signal to Noise Ratio}
\newacronym{son}{SON}{Self-Organizing Network}
\newacronym{ss}{SS}{Synchronization Signal}
\newacronym{srs}{SRS}{Sounding Reference Signal}
\newacronym{sss}{SSS}{Secondary Synchronization Signal}
\newacronym{tb}{TB}{Transport Block}
\newacronym{tcp}{TCP}{Transmission Control Protocol}
\newacronym{tdd}{TDD}{Time Division Duplexing}
\newacronym{tdma}{TDMA}{Time Division Multiple Access}
\newacronym{tfl}{TfL}{Transport for London}
\newacronym{tm}{TM}{Transparent Mode}
\newacronym{trp}{TRP}{Transmitter Receiver Pair}
\newacronym{tti}{TTI}{Transmission Time Interval}
\newacronym{ttt}{TTT}{Time-to-Trigger}
\newacronym{tx}{TX}{Transmitter}
\newacronym{ue}{UE}{User Equipment}
\newacronym{ul}{UL}{Uplink}
\newacronym{uml}{UML}{Unified Modeling Language}
\newacronym{um}{UM}{Unacknowledged Mode}
\newacronym{utc}{UTC}{Urban Traffic Control}
\newacronym{vm}{VM}{Virtual Machine}
\newacronym{rsrq}{RSRQ}{Reference Signal Received Quality}
\newacronym{rssi}{RSSI}{Received Signal Strength Indicator}
\newacronym{crs}{CRS}{Cell Reference Signal}
\newacronym{nsa}{NSA}{Non Stand Alone}
\newacronym{mrdc}{MR-DC}{Multi \gls{rat} \gls{dc}}
\newacronym{endc}{EN-DC}{E-UTRAN-\gls{nr} \gls{dc}}
\newacronym{5gc}{5GC}{5G Core}
\newacronym{si}{SI}{Study Item}
\newacronym{iab}{IAB}{Integrated Access and Backhaul}
\newacronym{wf}{WF}{Wired-first}
\newacronym{hqf}{HQF}{Highest-quality-first}
\newacronym{pa}{PA}{Position-aware}
\newacronym{mlr}{MLR}{Maximum-local-rate}
\newacronym{wbf}{WBF}{Wired Bias Function}
\newacronym{mib}{MIB}{Master Information Block}
\newacronym{sib}{SIB}{Secondary Information Block}
\newacronym{kpi}{KPI}{Key Performance Indicator}
\newacronym[plural=PPPs,firstplural=Poisson Point Processes (PPPs)]{ppp}{PPP}{Poisson Point Process}
\newacronym{vr}{VR}{Virtual Reality}
\newacronym{scm}{SCM}{Spatial Channel Model}